\documentclass{PoS}

\title{Exclusive production observed at the CMS experiment}

\ShortTitle{Exclusive production at CMS}

\author{\speaker{Ruchi Chudasama (for the CMS Collaboration)}\\
        Nuclear Physics Division \\
        Bhabha Atomic Research Center\\
        Mumbai, India\\
        E-mail: \email{ruchi.physics@gmail.com}}

\abstract{Exclusive W$^{+}$W$^{-}$ pair production in photon-photon collisions during the pp runs at 7 and 8 TeV are 
observed and used to put constraints on the Anomalous Quartic Gauge Couplings. During the proton lead collisions in
photon-induced vector meson production is observed via the decay of upsilon into two muons. The slope of the 
squared p$_{\rm T}$ distribution is measured to determine the size of the production region.}

\FullConference{XXIV International Workshop on Deep-Inelastic Scattering and Related Subjects\\
		11-15 April, 2016\\
		DESY Hamburg, Germany}

\begin{document}

\section{Introduction}
The CMS detector provides a very wide range of opportunities to study high-energy photon-induced interactions with proton and/or ion beams, 
due to the high energy and large integrated luminosities available at the LHC. Different exclusive particle 
production processes at high energies have been studied~\cite{ref1,ref2,ref3}. 
The very recent measurement of W pairs in the isolated electron-muon final-state have been used to extract the limit on anomalous quartic gauge coupling~\cite{ref2,ref3}. In addition, with the exclusive production of vector meson, nuclear gluon shadowing~\cite{ref4} and parton distribution function at very low x can be studied~\cite{ref9}.  

\section{Exclusive upsilon photoproduction in pPb collisions at 5.02 TeV}

\begin{figure*}[hbtp]
\begin{center}
\includegraphics[width=0.35\textwidth]{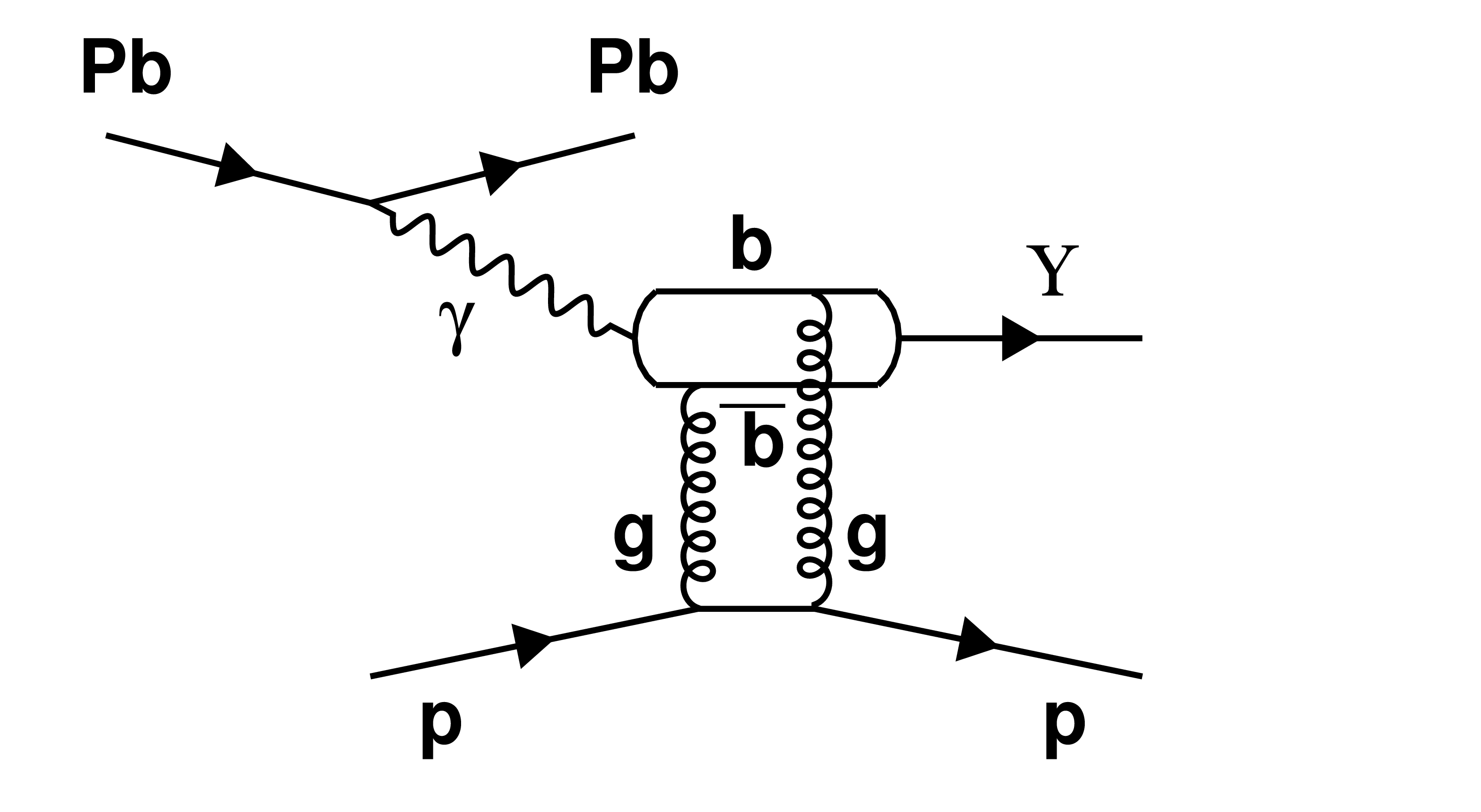}
\includegraphics[width=0.35\textwidth]{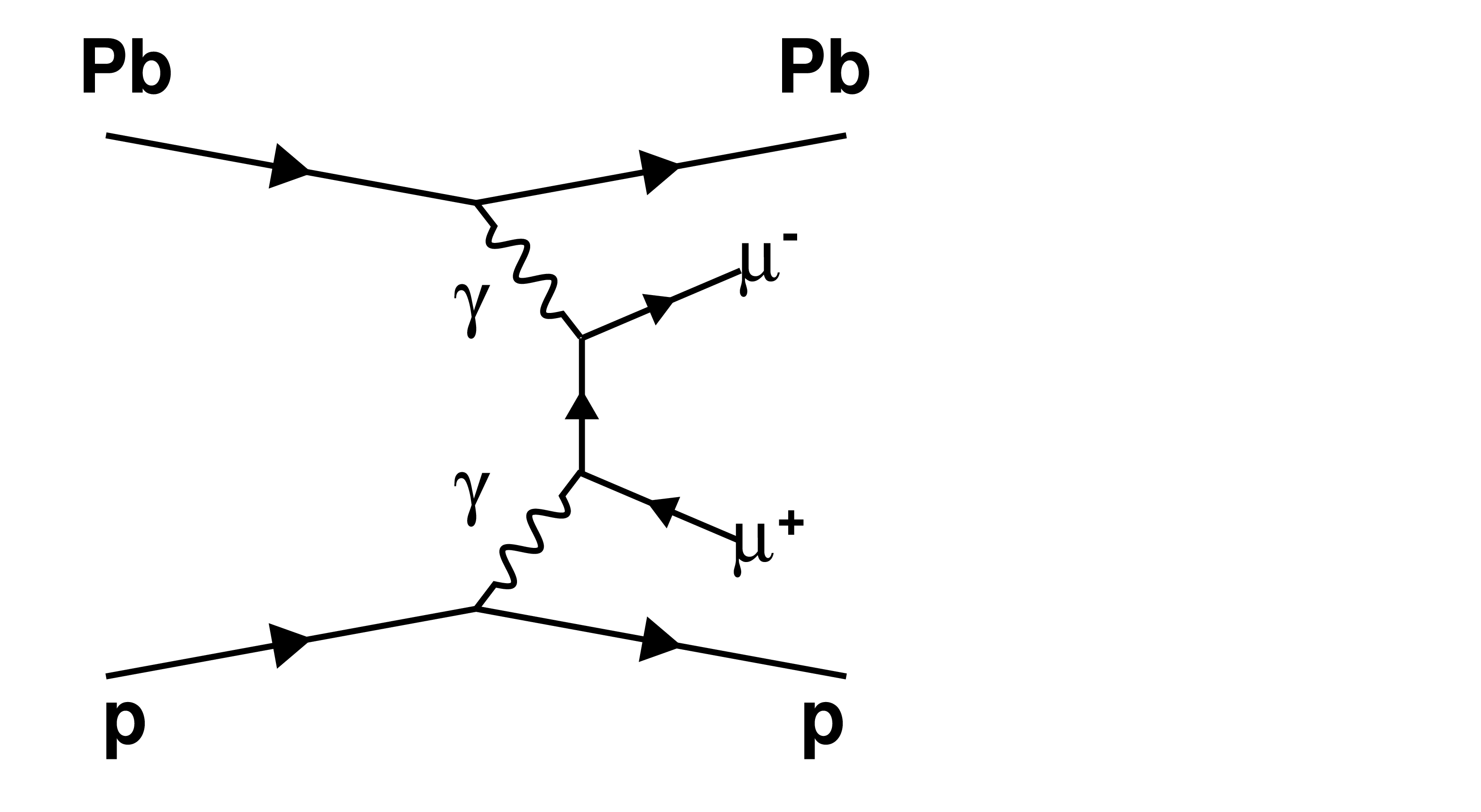}
\caption{Diagrams representing exclusive $\Upsilon$ photoproduction (left), and exclusive dimuon
QED continuum (right) in pPb collisions~\cite{ref9}.}
\label{fig:feynman}
\end{center}
\end{figure*}

Exclusive photoproduction of heavy vector mesons (Fig.~\ref{fig:feynman},left) at very high photon-proton 
center-of-mass energies $(W_{\gamma p})$ can be studied in ultraperipheral collisions (UPC)
of protons (ions). Recently, CMS, ALICE ~\cite{ref5} and LHCb~\cite{ref6} presented their measurements 
of exclusive heavy vector meson photoproduction at the LHC.
Since the process occurs through
$\gamma p$ or $\gamma Pb$ interaction via the exchange of two-gluons with
no net color transfer and thus, at leading order (LO),  the cross section  is proportional
to the  square of the gluon density in the target proton or ion.
It provides a valuable probe of the gluon density at the small 
 momentum fraction $x$ which is kinematically related to $W_{\gamma p}$
($x=(M_{\Upsilon} /W_{\gamma p})^2$).
 
The  exclusive photoproduction of $\Upsilon$(1S, 2S, 3S) measurement has been measured 
in their dimuon decay channel in ultraperipheral collisions 
of protons and heavy ions (pPb) with the CMS experiment at 
$\sqrt {s_{{\rm NN}}} = 5.02$ TeV for an integrated 
luminosity of $L_{{\rm int}} = 33$ nb$^{-1}$.
The photoproduction cross section for $\Upsilon$$(nS)$ was measured
 as a function of $W_{\gamma p}$ in the range $91< W_{\gamma p} < 826$ GeV 
which corresponds to the rapidity of
 the $\Upsilon$ meson in the range $|y| < 2.2 $ and $x$ values are of the order $x\sim 10^{-4}$ to $x\sim 1.3\cdot 10^{-2}$.
 The dependence of the elastic $\Upsilon$ photoproduction cross section on the
squared $\Upsilon$ transverse momentum  
approximating the four-momentum transfer at the proton vertex ($|t|\approx p_{\rm T}^2$), 
 can be parametrized  with an exponential function $e^{-b|t|}$ at low
 values of $|t|$. The differential cross section $d\sigma/dt$, has been measured in the range
 $|t| < 1.0$ (GeV/c)$^{2}$ and  the b-slope parameter was estimated.

The UPC events were selected by applying dedicated HLT trigger which
selects at least one muon in each event and at least one to six tracks. 
To select the exclusive $\Upsilon$(nS) events offline, two muon tracks originating from the same
primary vertex in each event were used. The muons were selected with $p_{\rm T} > 3.3$ GeV 
and pseudorapidity $|\eta|< 2.2$, in order to have high muon finding efficiency.
The $p_{\rm T}$ of the muon pair was selected between $0.1$ to $1$ GeV.
The lower cut on muon pair reduces the contamination from elastic QED background (Fig.~\ref{fig:feynman},right) and 
higher cut on muon pair reduces the contamination from inelastic background 
(proton dissociation, inclusive $\Upsilon$, Drell-Yan). The rapidity of muon pair is restricted to |y| $< 2.2$. 

The dominant background contribution to exclusive $\Upsilon$ signal comes from QED,  
$\gamma \gamma \rightarrow \mu^{+}\mu^{-}$, which was estimated by {\textsc{Starlight}}. 
The absolute prediction of QED was checked by comparing the data between invariant mass region 8--9.12 and 10.64--12 GeV  
for dimuon $p_{\rm T} < 0.15$ GeV to the simulation. The contribution of non-exclusive background 
(inclusive $\Upsilon$, Drell-Yan and proton dissociation) was estimated by a data-driven 
method by selecting events with more than 2 tracks. This template was normalized to two muon track sample in the region 
of dimuon $p_{T} > 1.5$ GeV. Additional background in this analysis originates from a small contribution of 
exclusive $\gamma {\rm Pb} \rightarrow$ $\Upsilon$ ${\rm Pb}$ events. The fraction of these events in the total
number of exclusive $\Upsilon$ events was estimated using the reweighted \textsc{Starlight} $\Upsilon$ MC sample.
These backgrounds were subtracted from data to get the exclusive signal. 

The background subtracted |t| and y distributions were used to measure the b parameter and estimate the exclusive 
$\Upsilon$ photoproduction cross-section as a function of $W_{\gamma p}$, respectively. The distributions were first
unfolded to the region $0.01<|t|<1$~GeV$^2$, $|y|<2.2$, and muon $p_{\rm T}^{\mu}>3.3$~GeV, using the 
iterative Bayesian unfolding technique and it's further extraolated to transverse momenta zero by acceptance correction factor. 

\begin{figure*}[t]
\begin{center}
\includegraphics[width=0.60\textwidth]{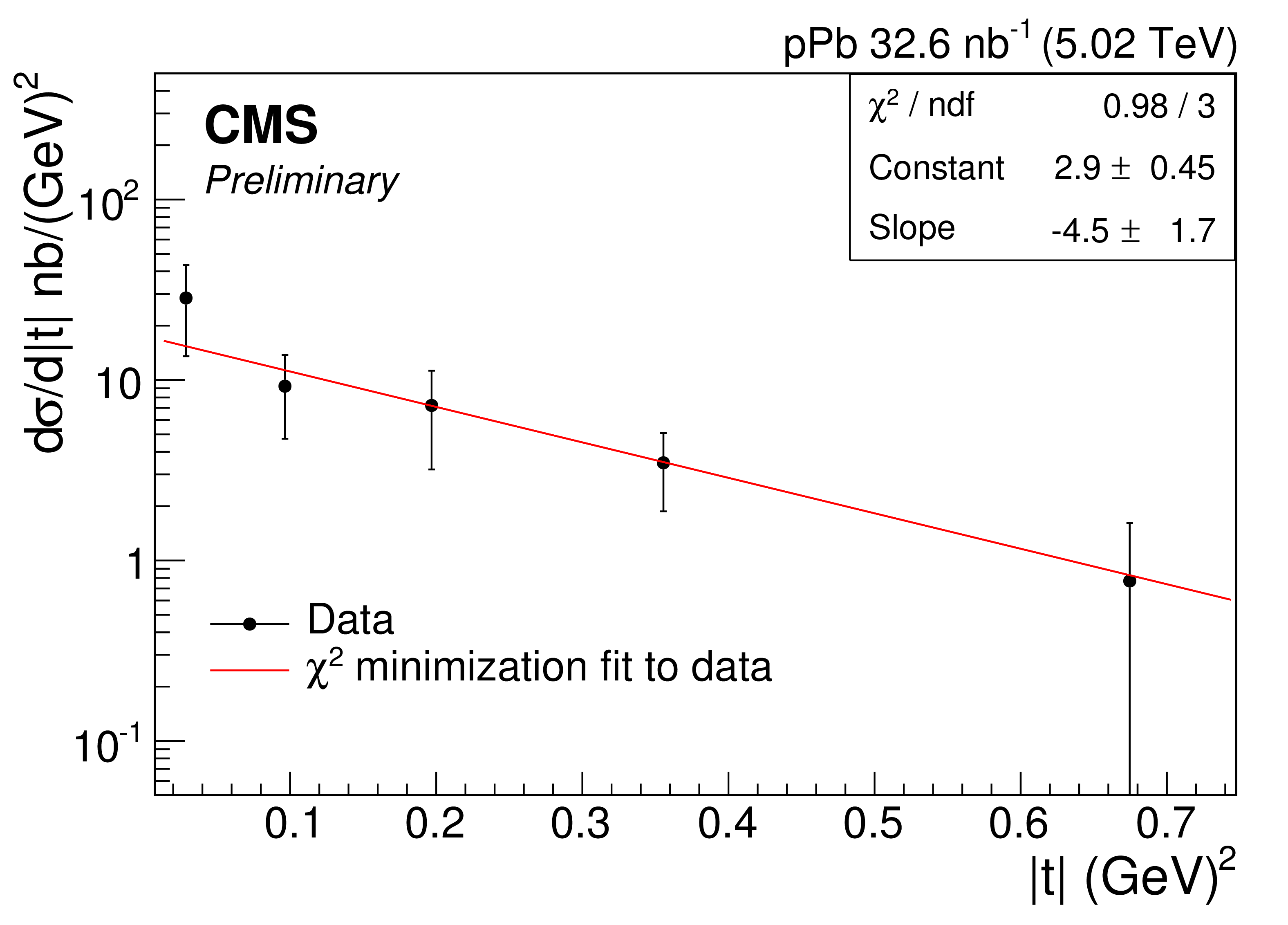}
\caption{Differential $\Upsilon$ photoproduction cross section as a function
of $|t|$ measured in pPb collisions at $\sqrt {s_{{\rm NN}}} = 5.02$ TeV in the dimuon rapidity
region $|y| < 2.2$. The solid line represents
the result of a fit with an exponential function $N e^{-b|t|}$~\cite{ref9}.}
\label{fig:pt2fit}
\end{center}
\end{figure*}

The differential $d\sigma/dt$ cross section is extracted for the combined three $\Upsilon$(nS) states as shown in Fig. ~\ref{fig:pt2fit},
according to 
\begin{equation}
\frac{{\rm d}\sigma_{\Upsilon}}{{\rm d}|t|}=\frac{N^{\Upsilon{\rm (nS)}}}{\mathcal{L} \times \Delta |t|}~,
\end{equation}
where $|t|$ is approximated by the dimuon transverse momentum squared $p_\mathrm{T}^2$, $N^{\Upsilon{\rm (nS)}}$
denotes the background-subtracted, unfolded and acceptance-corrected number of signal events in each $|t|$ bin, 
$\mathcal{L}$ is the integrated luminosity, and $\Delta |t|$ is the width of each $|t|$ bin.  
The cross section is fitted with an exponential function $N~e^{ -b|t|}$ in the region
$0.01<|t|<1.0$ GeV$^2$, using an unbinned $\chi^2$ minimization method. 
A value of $b=4.5 \pm 1.7$ (stat) $\pm$ $0.6$ (syst) GeV$^{-2}$ 
is extracted from the fit. This result is in
agreement with the value $b=4.3^{+2.0}_{-1.3}$ (stat) measured by 
the ZEUS experiment~\cite{ref13} for the
photon-proton center-of-mass energy $60<W_{\gamma {\rm p}}<220$~GeV.
The measured value of $b$ is also
consistent with the predictions based on pQCD models~\cite{ref14}.  

The differential $\Upsilon$(1S) photoproduction cross section ${\rm d}\sigma/{\rm d}y$ is extracted in four
bins of dimuon rapidity according to
\begin{equation}
\label{eq:cross_dsigmady}
\frac{d\sigma_{\Upsilon({\rm 1S})}}{dy}=\frac{f_{\Upsilon({\rm 1S})}}{\mathcal{B}(1+f_{\rm FD})}
\frac{N^{\Upsilon{\rm (nS)}}}{\mathcal{L} \times \Delta y}~,
\end{equation}
where $N^{\Upsilon{\rm (nS)}}$ denotes the background-subtracted, unfolded and acceptance-corrected number of
signal events in each rapidity bin. The factor $f_{\Upsilon({\rm 1S)}}$ describes the ratio of $\Upsilon${\rm
  (1S)} to $\Upsilon${\rm (nS)} events, $f_{\rm FD}$ is the feed-down contribution to the $\Upsilon${\rm (1S)}
events originating from the $\Upsilon$(2S)$\rightarrow \Upsilon({\rm 1S}) + X$ decays (where
$X=\pi^{+}\pi^{-}$ or $\pi^{0}\pi^{0}$), $\mathcal{B} = (2.48\pm 0.05)\%$ is the branching
ratio for muonic $\Upsilon(1S)$ decays, and $\Delta y$ is the width of the $y$ bin.

The $f_{\Upsilon({\rm 1S)}}$ fraction is used from the results of the inclusive $\Upsilon$ analysis~\cite{ref10}.
The feed-down contribution of $\Upsilon$(2S) decaying to $\Upsilon(1S) + \pi^{+}\pi^{-}$ and $\Upsilon(1S) + \pi^{0}\pi^{0}$ 
was estimated as $15$\%  from the \textsc{Starlight}. The contribution from feed-down of exclusive $\chi_b$ states was neglected,
as these double-pomeron processes are expected to be comparatively much suppressed in proton-nucleus collisions~\cite{ref11,ref12}.
\begin{figure*}[t]
\begin{center}
\includegraphics[width=0.60\textwidth]{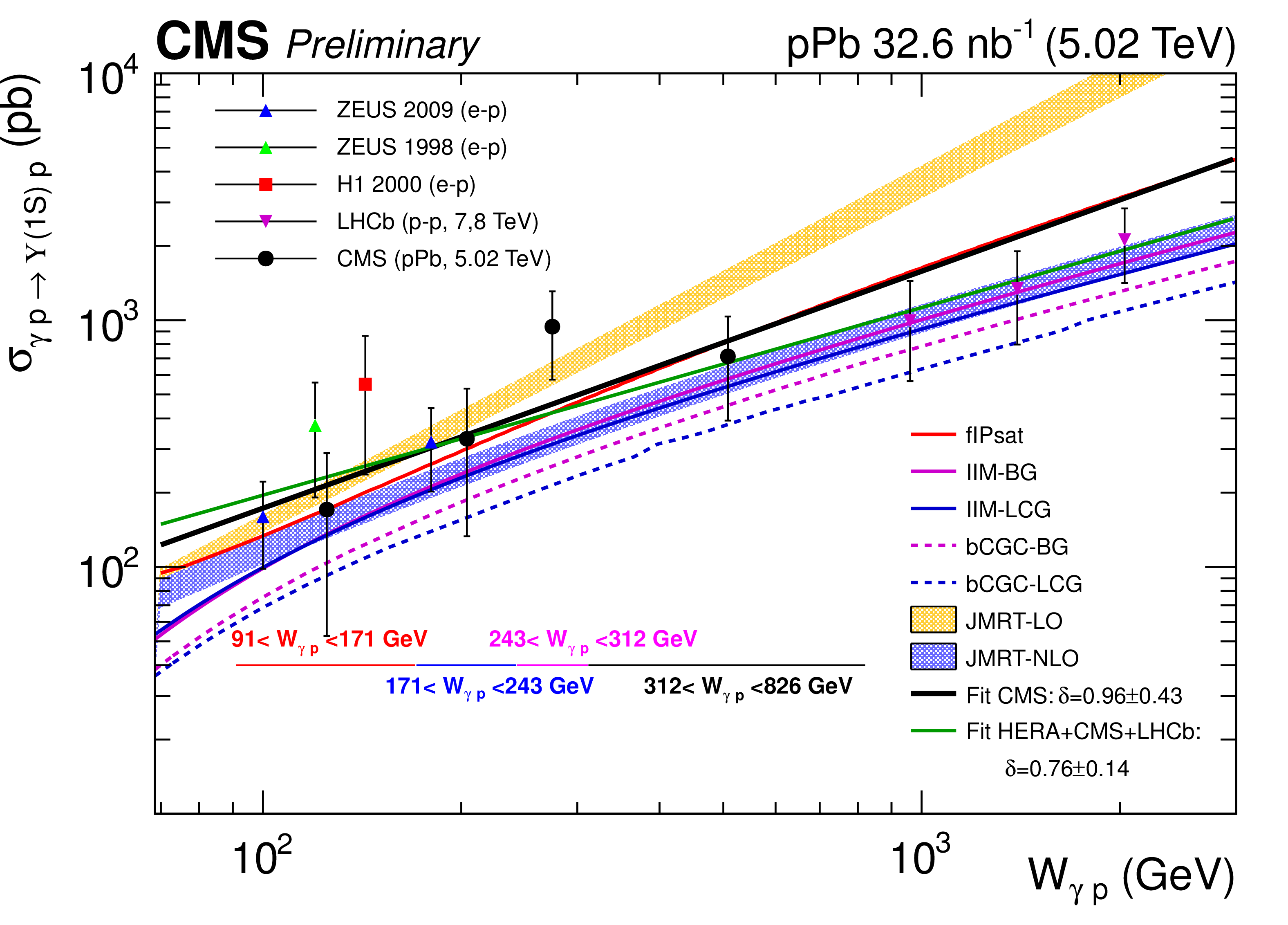}
\caption{Cross section for exclusive $\Upsilon$(1S) photoproduction, $\gamma p \rightarrow \Upsilon (1S) p$ as a function of photon-proton center-of-mass energy, $W_{\gamma p}$~\cite{ref9}.}
\label{fig:wgp}
\end{center}
\end{figure*}

The exclusive $\Upsilon$(1S) photoproduction cross section as a 
function of $W_{\gamma p}$ as shown in Figure~\ref{fig:wgp}, is obtained by using, 
\begin{equation}
\sigma_{\gamma p \rightarrow \Upsilon(1S)p}(W_{\gamma p}^{2}) = \frac{1}{\Phi}\frac{d\sigma_{\Upsilon(1S)}}{dy},
\label{eq:photo_cross}
\end{equation}
where $\Phi$ is the  photon flux evaluated at the mean of the rapidity bin, 
estimated from {\textsc{Starlight}}. The CMS data are plotted together with the previous
measurements from H1~\cite{ref7}, ZEUS~\cite{ref8} and 
LHCb ~\cite{ref2} data. It is also compared with different theoretical 
predictions of the JMRT model~\cite{ref14},
factorized IPsat model~\cite{ref15,ref16}, IIM~\cite{ref17,ref18} and 
bCGC model~\cite{ref19}. As $\sigma(W_{\gamma p})$ is proportional to
the square of the gluon PDF of the proton and the gluon distribution
at low Bjorken $x$ is well described by a power law,  
the cross section will also follow a power law. Any deviation 
from such trend would indicate different behavior of gluon density function. 
We fit a power law $A\times (W/400)^\delta$ 
with CMS data alone which gives $\delta=0.96\pm 0.43$ and $A=655\pm 196$ 
and is shown by the black solid line. 
The extracted $\delta$ value is comparable to the value
$\delta=1.2 \pm 0.8$, obtained by ZEUS~\cite{ref4}.  

\section{Measurement of W$^{+}$W$^{-}$ pair production in pp collisions at 7 and 8 TeV}
\begin{figure*}[hbtp]
\begin{center}
\includegraphics[width=0.30\textwidth]{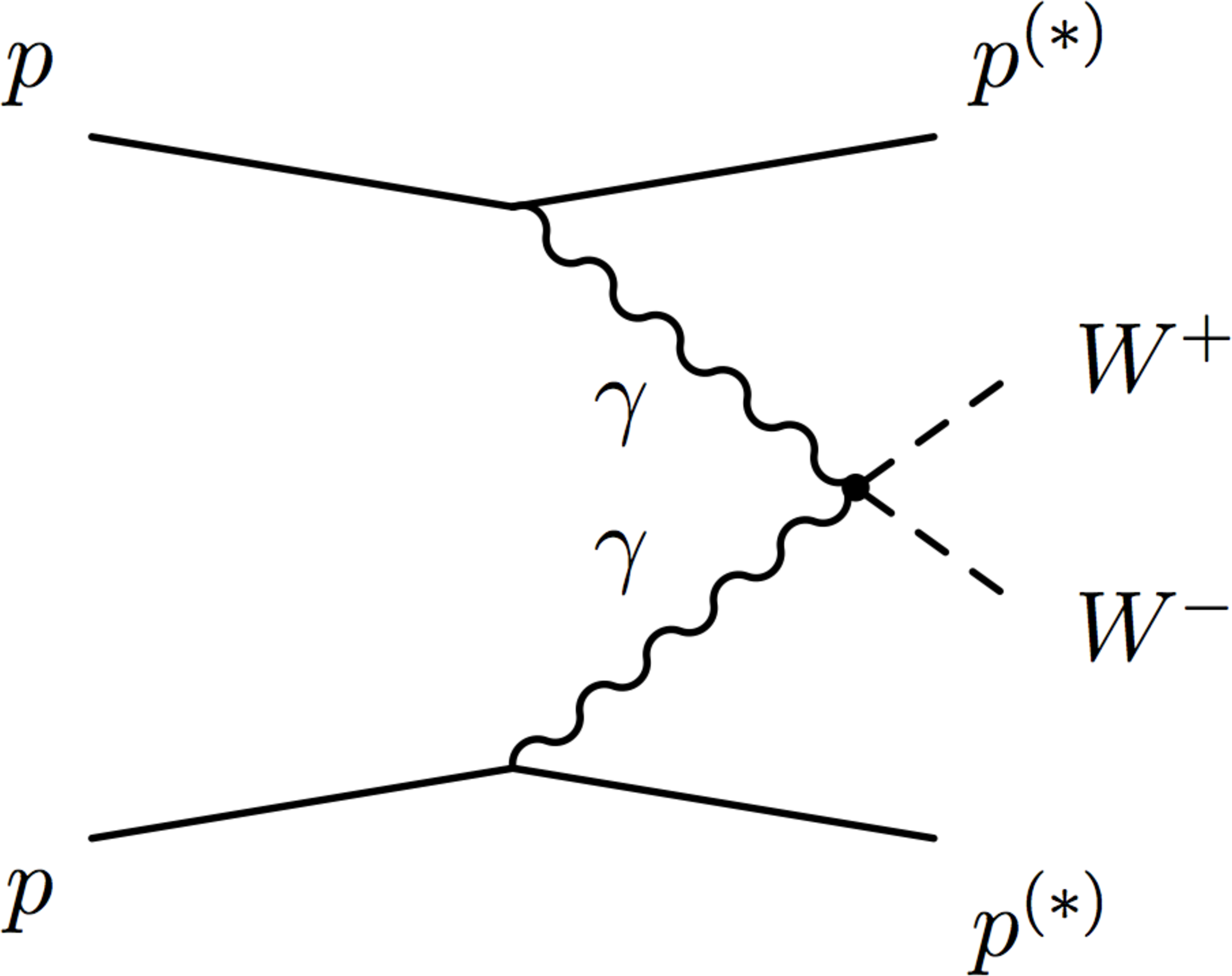}
\includegraphics[width=0.30\textwidth]{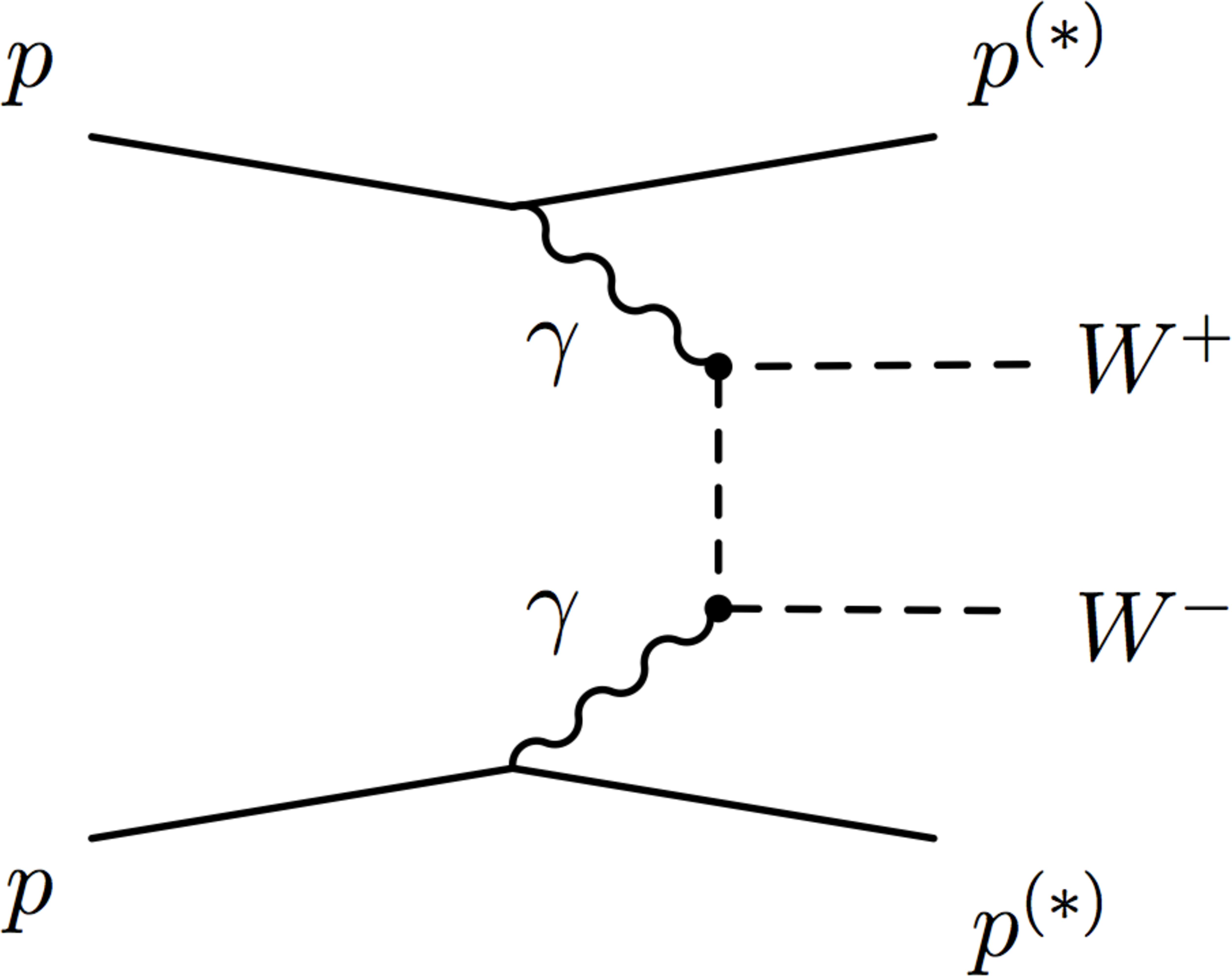}
\includegraphics[width=0.30\textwidth]{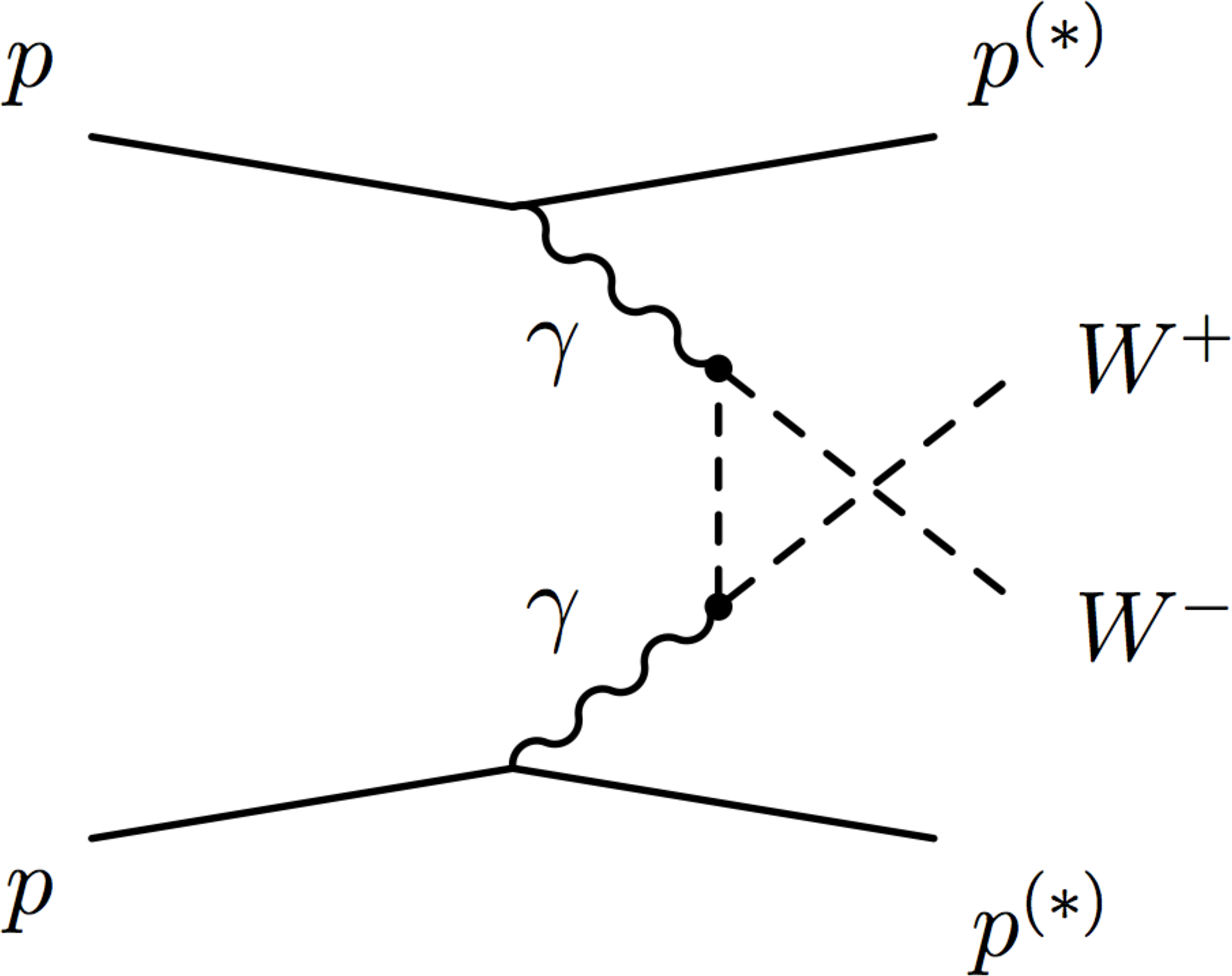}
\caption{Quartic (left), t-channel (center), and u-channel (right) diagrams contributing to the
$\gamma \gamma \rightarrow W^{+} W^{-}$ process at leading order in the SM~\cite{ref3}.}
\label{fig:feynman2}
\end{center}
\end{figure*}

High energy photon interaction at LHC provide a unique opportunity to study exclusive 
production of W pairs. At leading order, quartic , t-channel and u-channel processes contributes
to $\gamma \gamma \rightarrow W^{+} W^{-}$ production (Fig.~\ref{fig:feynman2}).  
The measurement of the quartic WW$\gamma\gamma$ coupling would provide an opportunity to look 
for any deviations from Standard Model (SM) and search for physics beyond SM. 
A genuine anomalous quartic gauge coupling (AQGC) is introduced via an effective lagrangian with two additional
dimension-6 terms containing the parameters a$_{0}^{W}$ and a$_{C}^{W}$. With the discovery of a light Higgs boson~\cite{ref22,ref23,ref24}
a linear realization of the SU(2)$\times$U(1) symmetry of the SM, spontaneously broken by the Higgs mechanism, is possible. 
Thus, the lowest order operators, where new physics may cause deviations in the quartic gauge boson
couplings alone, are of dimension 8. By assuming the the anomalous WWZ$\gamma$ vertex vanishes, a direct relationship
between the dimension-8 and dimension-6 couplings can be recovered~\cite{ref25}. In both dimension-6 and dimension-8 scenarios,
the $\gamma \gamma \rightarrow W^{+} W^{-}$ cross section increase quadratically with energy, therefore a dipole form 
factor is introduced to preserve unitarity.

CMS has carried out a measurement of exclusive and quasi-exclusive $\gamma \gamma \rightarrow W^{+} W^{-}$ production, 
via $pp \rightarrow p^{(*)}W^{+} W^{-}p^{(*)} \rightarrow p^{(*)} \mu^{\pm}e^{\mp} $ 
at 7 and 8 TeV , corresponding to luminosities of 5.5 fb$^{-1}$ and 19.7 fb$^{-1}$, respectively. 
The production of W pairs was measured in $\mu^{\pm}e^{\mp} $ final state, since, $W^{+} W^{-} \rightarrow \mu^{+}\mu^{-}$ 
or $W^{+} W^{-} \rightarrow e^{+}e^{-}$ would be dominated by Drell-Yan events and $\gamma \gamma \rightarrow l^{+}l^{-}$ production. 
The $\mu^{\pm}e^{\mp} $  events were extracted by skimming the dataset with dedicated HLT trigger which selects two 
leptons with transverse momentum $p_{T} > 17(8)$ GeV for the leading (subleading) lepton. 
Offline, the events with the opposite-charge electron-muon pair originating from a common primary vertex that has no 
additional tracks associated with it were selected to remove the underlying event activity. 
The events with transverse momentum of the pair p$_{T}$($\mu^{\pm}e^{\mp}$) $>$ 30 GeV were selected to suppress backgrounds
from $\tau^{+} \tau^{-}$ production, including the exclusive and quasi-exclusive $\gamma \gamma \rightarrow \tau^{+} \tau^{-}$ processes.

Simulations show that in high-mass $\gamma\gamma$ interactions one or both of the protons dissociate, which
may result in events being rejected by the veto on extra tracks. To estimate this effect from data, a
sample is selected where the dilepton invariant mass is greater than 160 GeV so that W$^{+}$W$^{-}$ pairs
can be produced on shell. The ratio of the observed number of events to the calculated number
of elastic pp $\rightarrow$ p$l^{+}l^{-}$p events is used as a scale factor to calculate from the predicted elastic
pp $\rightarrow$ pW$^{+}$W$^{-}$p events the total number of pp $\rightarrow$ p$^{*}$ W$^{+}$W$^{-}$ p$^{*}$ to be expected when including
proton dissociation. The numerical value of the scale factor thus obtained is F~=~4.10$\pm$0.43.

\begin{figure*}[hbtp]
\begin{center}
\includegraphics[width=0.40\textwidth]{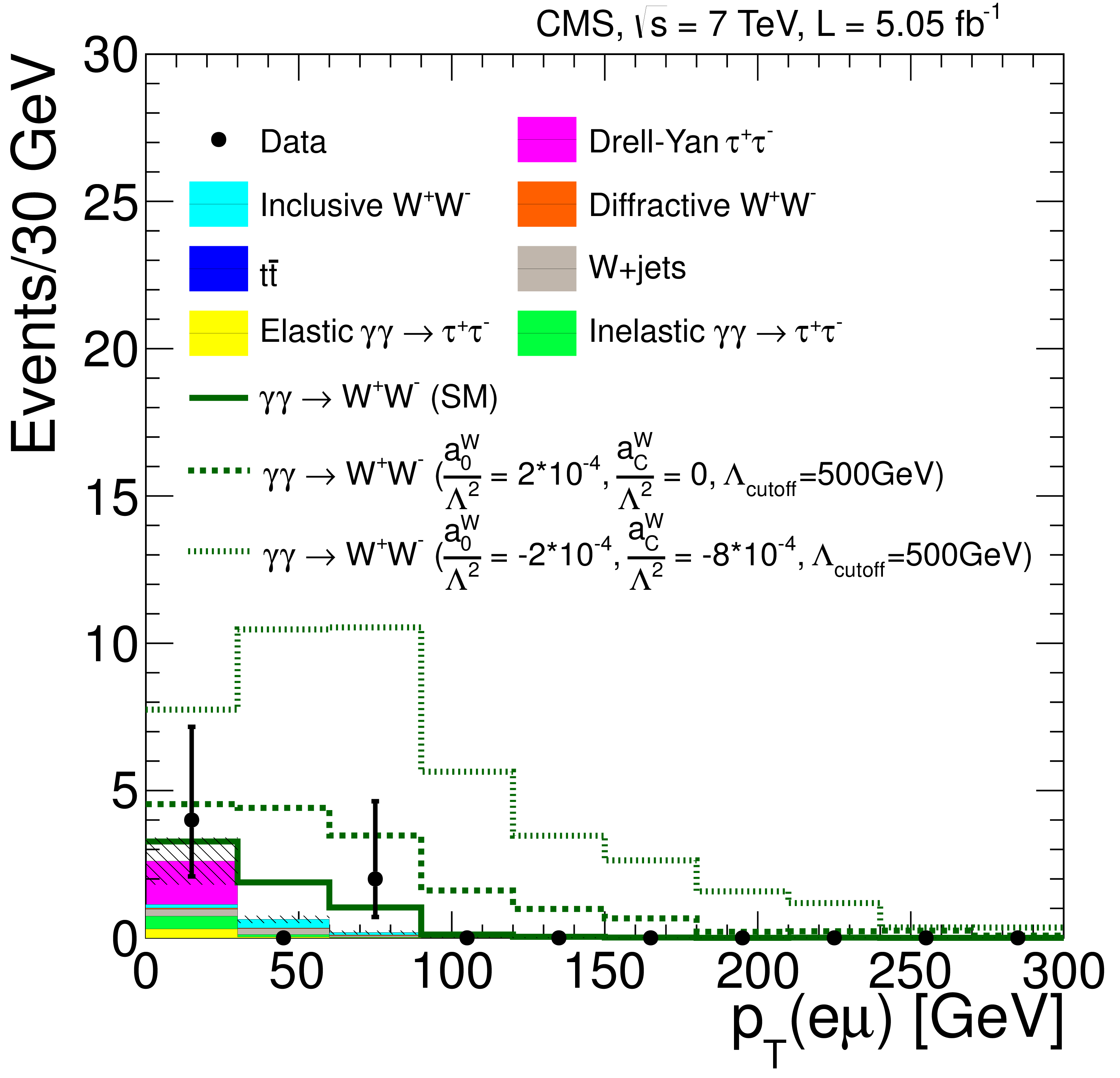}
\includegraphics[width=0.50\textwidth]{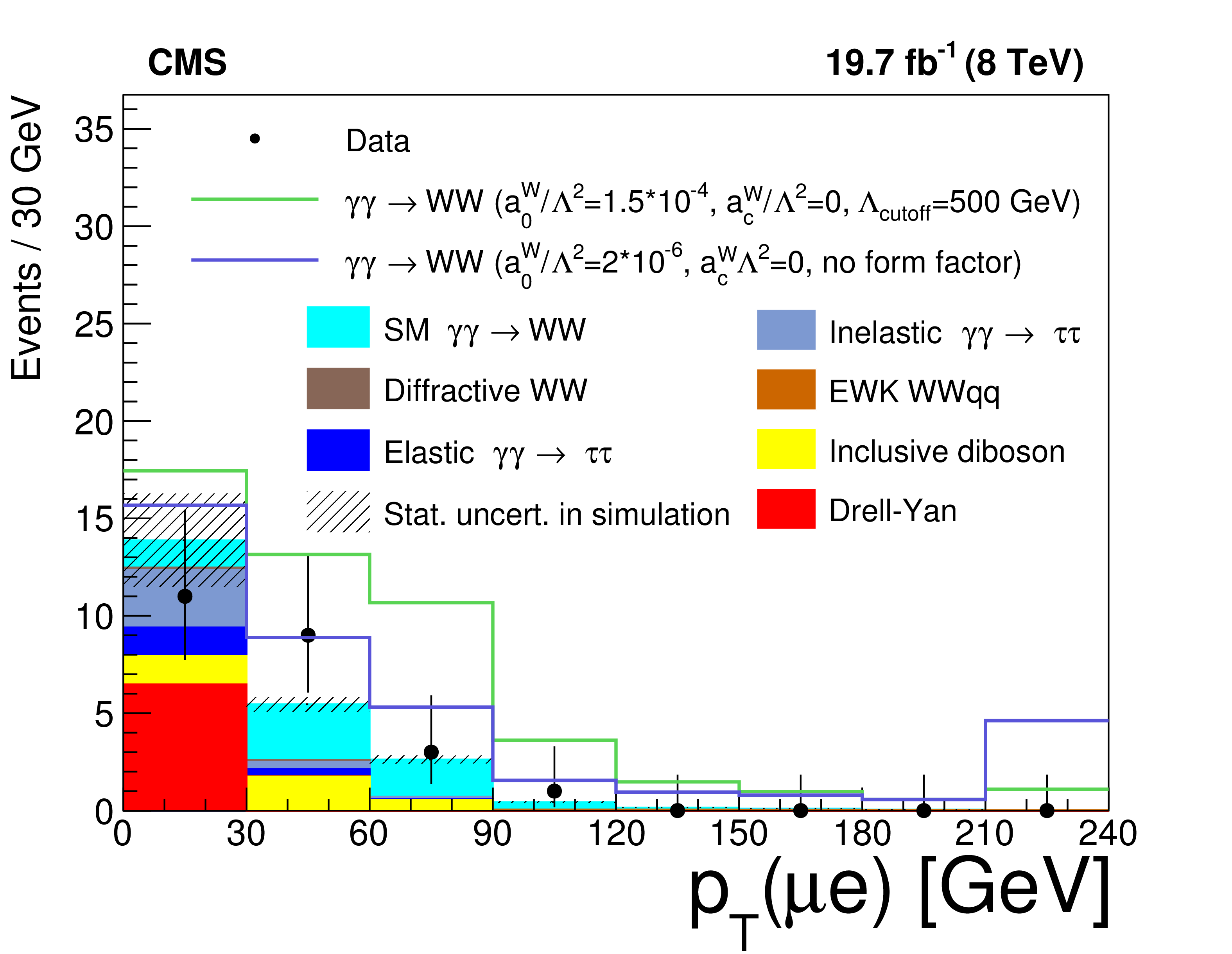}
\caption{ The p$_{T}$($\mu^{\pm}e^{\mp}$) distribution for events with zero extra tracks at 7 TeV (left)~\cite{ref2} and at 8 TeV(right)~\cite{ref3}.}
\label{fig:ptemu}
\end{center}
\end{figure*}

Fig.~\ref{fig:ptemu} shows the p$_{T}$($\mu^{\pm}e^{\mp}$) distribution for events passing all other selection requirements. 
In the signal region with no additional tracks and p$_{T}$($\mu^{\pm}e^{\mp}$) $>$ 30 GeV, 2 events are observed at
7 TeV compared to the expectation of 2.2$\pm$0.4 signal events and 0.84$\pm$0.15 background events, corresponding to an observed 
(expected) significance of 0.8$\sigma$ (1.8$\sigma$). While 13 events are observed at 8 TeV compared to the expectation of 5.3$\pm$0.7 
signal events and 3.9$\pm$0.6 background events, corresponds to a mean expected signal significance of 2.1 $\sigma$. We combine 
the 7 and 8 TeV results, treating all systematic uncertainties as fully uncorrelated between the two measurements, the resulting observed
(expected) significance for the 7 and 8 TeV combination is 3.4$\sigma$ (2.8$\sigma$), constituting evidence for
$\gamma \gamma \rightarrow W^{+} W^{-}$ production in proton-proton collisions at the LHC. Interpreting the 8 TeV results as a cross section 
multiplied by branching fraction to $\mu^{\pm}e^{\mp}$) final states, corrected for all experimental efficiencies and extrapolated to the
full space, yields :
\begin{equation}
\sigma(\rm pp \rightarrow p^{(*)} W^{+}W^{-} p^{(*)} \rightarrow p^{(*)} \mu^{\pm}e^{\mp} p^{(*)}) = 11.9^{+5.6}_{-4.5} fb
\label{eq:cross}
\end{equation}
The corresponding 95\% confidence level (CL) upper limit obtained from the 7 TeV data was $<$10.6 fb, with a central value of 2.2$^{+3.3}_{-2.0}$ fb. 

\begin{figure*}[hbtp]
\begin{center}
\includegraphics[width=0.50\textwidth]{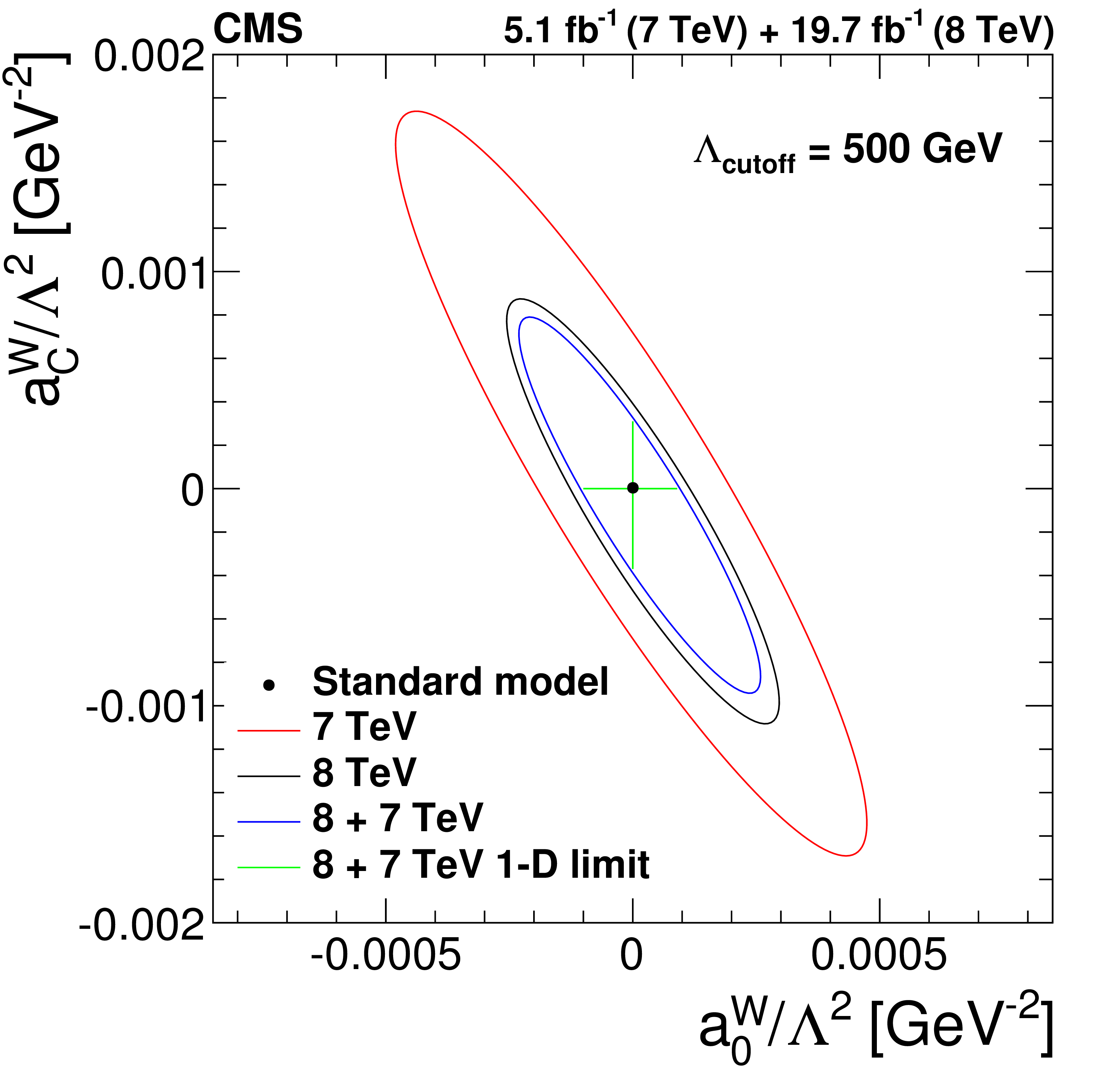}
\caption{Excluded values of the anomalous coupling parameters a$_{0}^{W}/\wedge^{2}$ and a$_{C}^{W}/\wedge^{2}$ with $\wedge_{cutoff}=500$ GeV. 
The areas outside the solid contours are excluded by each measurement at 95\% CL~\cite{ref3}.}
\label{fig:aqgc}
\end{center}
\end{figure*}

The transverse momentum p$_{T}$($\mu^{\pm}e^{\mp}$) distribution was also used to search for sign of anomalous quartic gauge couplings. 
p$_{T}$($\mu^{\pm}e^{\mp}$) $>$ 100 GeV is used at 7 TeV, while two bins, with boundaries p$_{T}$($\mu^{\pm}e^{\mp}$) = 30-130 GeV and 
p$_{T}$($\mu^{\pm}e^{\mp}$) $>$ 130 GeV, are used to set the limit on aQGC. Fig.~\ref{fig:aqgc} shows the excluded values of the anomalous coupling
 parameters a$_{0}^{W}/\wedge^{2}$ and a$_{C}^{W}/\wedge^{2}$ with $\wedge_{cutoff}=500$ GeV. The exclusion regions are shown at 7TeV (outer contour), 
8TeV (middle contour), and the 7+8 TeV combination (innermost contour). The areas outside the solid contours are excluded by each measurement at 95\% CL.
The cross indicates the one-dimensional limits obtained for each parameter from the 7 and 8 TeV combination, with the other parameter fixed to zero, 
more details can be found in ~\cite{ref3}.

\section{Summary}
We reported the first measurement of the exclusive photoproduction of  $\Upsilon$(1S, 2S, 3S) mesons in the
$\mu^{+}\mu^{-}$ decay modes in ultraperipheral pPb collisions at $\sqrt {s_{{\rm NN}}} = 5.02$ TeV, 
corresponding to an integrated luminosity of $33$~nb$^{-1}$. The exclusive photoproduction cross sections have
 been measured as a function of the photon-proton center of mass
energy, bridging a previously unexplored region between HERA and LHCb measurements.
Our data are compatible with a power law dependence of $\sigma(W_{\gamma p})$, disfavouring faster rising
predictions  of  LO pQCD. The spectral slope b has been extracted, in agreement with earlier measurements. 
Results have been also presented for exclusive and quasi-exclusive $\gamma \gamma \rightarrow W^{+} W^{-}$ production in the 
$\mu^{\pm}e^{\mp}$ final state in pp collisions at $\sqrt {s_{{\rm NN}}}$ = 7 and 8 TeV, corresponding to an integrated 
luminosity of $5.5$~fb$^{-1}$ and $19.7$~fb$^{-1}$, respectively. The observed yields and kinematic distributions are 
consistent with the SM prediction, with a combined significance over the background-only hypothesis of 3.4 $\sigma$.
No significant deviations from the SM are observed in the p$_{T}$($\mu^{\pm}e^{\mp}$) distribution.


\begin{thebibliography}{99}
\bibitem{ref1} CMS Collaboration, JHEP 01 (2012) 052  
\bibitem{ref2} CMS Collaboration, JHEP 07 (2013) 116 
\bibitem{ref3} CMS Collaboration, arXiv: 1604.04464 
\bibitem{ref4} CMS Collaboration, arXiv: 1605.06966   
\bibitem{ref5} ALICE Collaboration, Phys. Rev. Lett. 113 (2014) 232504. 
\bibitem{ref6} LHCb Collaboration, JHEP 09 (2015) 084, arXiv:1505.08139.  
\bibitem{ref7} H1 Collaboration, Eur. Phys. B 46 (2006) 585.
\bibitem{ref8} ZEUS Collaboration, Phys. Lett. B 680 (2009) 4.
\bibitem{ref9} CMS Collaboration, FSQ-13-009
\bibitem{ref10} CMS Collaboration, JHEP 04 (2014) 103.
\bibitem{ref11} A. J. Schramm and D. H. Reeves, Phys. Rev. D 55 (1997) 7312.
\bibitem{ref12} L. A. Harland-Lang, V. A. Khoze, M. G. Ryskin, and W. J. Stirling, Eur. Phys. J. C 69 (2010) 179.
\bibitem{ref13} ZEUS Collaboration, Phys. Lett. B 708 (2012) 14.
\bibitem{ref14} P. Jones, D. Martin, M. G. Ryskin, and T. Teubner, JHEP 11 (2013) 085.
\bibitem{ref15} T. Lappi and H. Mantysaari, Phys. Rev. C 83 (2011) 065202.
\bibitem{ref16} T. Lappi and H. Mantysaari, Phys. Rev. C 87.
\bibitem{ref17} G. Sampaio dos Santos and M. V. T. Machado, Phys. Rev. C 89 (2014) 025201,(2013) 032201. 
\bibitem{ref18} G. Sampaio dos Santos and M. V. T. Machado, J. Phys. G42 (2015) 105001.
\bibitem{ref19} V. P. Goncalves, B. D. Moreira, and F. S. Navarra, Phys. Lett. B 742 (2015) 172. 
\bibitem{ref20} O. J. P. Eboli, M. C. Gonzalez-Garcia and S. M. Lietti, Phys. Rev. D 69, 095005 (2004).
\bibitem{ref21} O. J. P. Eboli, M. C. Gonzalez-Garcia and J. K. Mizukoshi, Phys. Rev. D 74, 073005 (2006)
\bibitem{ref22} ATLAS Collaboration, Phys. Lett. B 716 (2012) 1
\bibitem{ref23} CMS Collaboration, Phys. Lett. B 716 (2012) 30
\bibitem{ref24} CMS Collaboration, JHEP 06 (2013) 081
\bibitem{ref25} CMS Collaboration, Phys. Rev. D 90 (2014) 032008

\end{thebibliography}
\end{document}